 \documentclass[smallabstract,smallcaptions]{dccpaper}

\usepackage{epsfig}
\usepackage{citesort}
\usepackage{amsmath}
\usepackage{amssymb}
\usepackage{color}
\usepackage{url}
\usepackage{amssymb}
\usepackage{multirow}
\usepackage{algorithm,algpseudocode}

\newlength{\figurewidth}
\newlength{\smallfigurewidth}

\begin{document}

\title
{\large
\textbf{An Entropy Coding Based on Binary Encoding for Mixed-Radix Digits}
}

\author{%
	Na Wang$^{\ast}$, Wei Yan$\dag$, Sian-Jheng Lin$^{\ddag}$, Yuliang Huang$^{\ddag}$\\[0.5em]
	{\small\begin{minipage}{\linewidth}\begin{center}
				\resizebox{\textwidth}{!}
				{
					\begin{tabular}{ccccc}
						$^{\ast}$University of Shanghai for &&$\dag$University of Science and && $^{\ddag}$ Theory Lab, Cenrtal Research  \\
						Science and Technology$\left(\text{USST}\right)$, && Technology of China$\left(\text{USTC}\right)$, &&
						Institute, 2012 Labs,\\
						Shanghai, China,  && Heifei, China,  && Huawei Technology Co. Ltd,\\
						\url{wna@usst.edu.cn} && \url{yan1993@mail.ustc.edu.cn} && \url{{lin.sian.jheng1,huangyuliang1}@huawei.com}
					\end{tabular}
				}
			\end{center}
	\end{minipage}}
}

\maketitle
\thispagestyle{empty}

\begin{abstract}
The necessity of radix conversion of numeric data is an indispensable component in any complete analysis of digital computation. In this paper, we propose a binary encoding for mixed-radix digits. Second, a variant of rANS coding based on this conversion is given, which supports parallel decoding. The simulations show that the proposed coding in serial mode has a higher throughput than the baseline (with the speed-up factor about 2$\times$) without loss of compression ratio, and it outperforms the existing 2-way interleaving implementation.
\end{abstract}

\section{Introduction}
Entropy coding~\cite{8416581,7786181} is a type of lossless coding to compress digital data by representing frequently occurring patterns with few bits and rarely occurring patterns with many bits. 
A number of well-known entropy encodings include Huffman coding~\cite{9105909}, arithmetic coding and asymmetric numeral systems (ANS)~\cite{duda2013asymmetric}. In particular, ANS is a new approach to entropy coding proposed by Jarek Duda in 2009, providing the speed comparable with Huffman coding, as well as the compression ratio similar to arithmetic coding. This leads that ANS is an alternative to entropy coding in many compressors, such as Facebook Zstandard and Google Draco 3D. There are several versions in ANS, and range ANS (rANS) is one of the major versions of ANS. Recently, the research~\cite{8849430,9478894} on it has drawn many academic interests since it is a more efficient way for data storage and transmission.

In this paper, we propose a variant of rANS coding based on a binary encoding for mixed-radix digits~(BEMR). BEMR provides the conversion between a set of mixed-radix digits and a binary stream, such as Szabo-Tanaka Mixed-Radix Conversion (MRC)~\cite{szaboresidue}, truncated binary encoding~\cite{mahapatra2011inverted} and Chen–Ho encoding~\cite{chen1975storage}. Among them, Chen–Ho encoding gives a method to encode three decimal digits into a $10$-bit codeword with simple Boolean operations only. However, Chen–Ho encoding only supports the conversion for base-$10$ digits. Further, truncated binary encoding provides a prefix coding used when the base is not a power of two. Nevertheless, its coding efficiency is lower, especially for small bases. In this paper, we propose a BEMR by applying the renormalization on radix conversion. Then, an entropy coding based on the proposed BEMR is presented. The contributions of this paper are summarized as follows.
\begin{enumerate}
	\item By applying the renormalization on radix conversion, a BEMR is proposed.
	\item Based on the proposed BEMR, a variant of rANS coding is given. Compared with traditional rANS, the proposed version supports parallel decoding.
	\item The simulations show that the proposed scheme in serial mode has a higher throughput than the baseline (with the speed-up factor about 2$\times$) without loss of compression ratio, and it outperforms the existing 2-way interleaving implementation.
\end{enumerate}

The remainder of this paper is organized as follows. Section~\ref{sec:2} lists the notations and related works. Section~\ref{sec:3} introduces the proposed BEMR and a variant of rANS coding. Section~\ref{sec:4} shows the simulation results. Finally, Section~\ref{sec:5} concludes this paper with an outlook on future work.

\section{Preliminaries}\label{sec:2}
\subsection{Notations}\label{sec:2.1}
Let $|X|_m$ denote the remainder of a number $X$ with respect to a modulus $m$. Let $\left [a,b \right ):=\left\{a,a+1,\dots ,b-1\right\}$.

For a double-ended queue (abbreviated to deque) with the feature of push and pop operations for both front and back, let the functions $push\_front()$, $push\_back()$, $pop\_front()$ and $pop\_back()$ represent the insertion and deletion from the front and back of deque, respectively.

Given a binary stream $E$, a deque is constructed accordingly. The operation
\begin{equation}
	Q\gets\mathrm{Deque}_E\left(i\right)
\end{equation}
pops $i$ bits $\left\{E_i\right\}_{i=0}^{i-1}$ from the front terminal position in the deque containing $E$, and the returned value forms an integer $Q=E_0+E_1\times 2+\dots + E_{i-1}\times 2^{i-1}$.

\begin{figure}[t]
	\centering
	\includegraphics[width=1.0\linewidth]{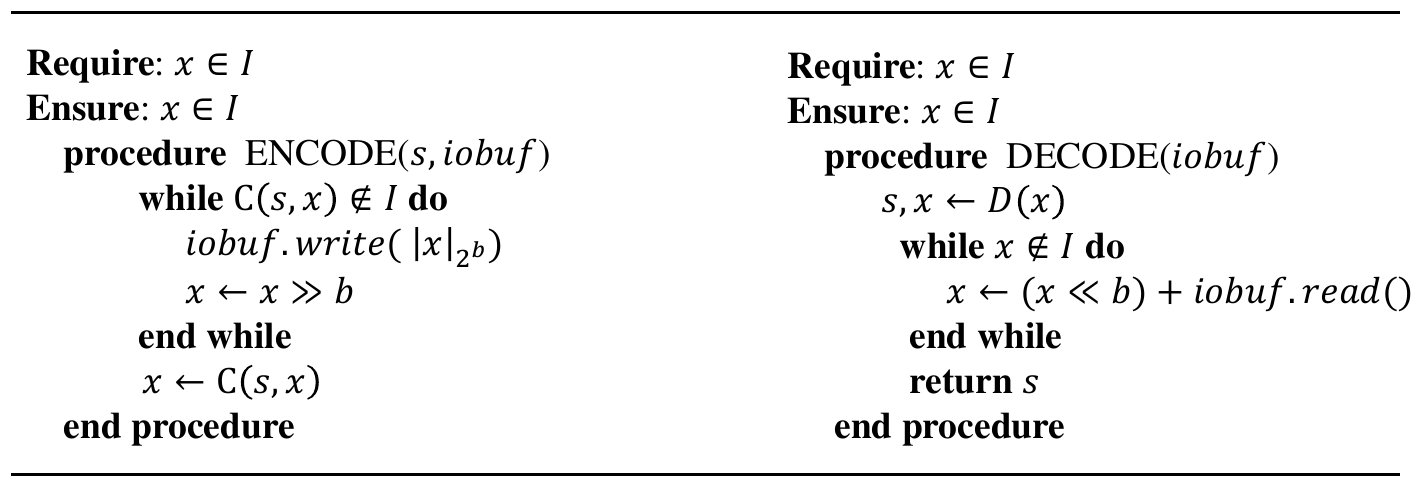}
	\caption{Streaming rANS encoder/decoder~\cite{giesen2014interleaved}.
	} 
	\label{rANS}   
\end{figure}

\subsection{rANS}\label{sec:2.3}
Given a finite symbol alphabet $\Sigma:=\left [0,z \right )$. Assuming that the frequency distribution $F= \left\{f_0,f_1,\cdots, f_{z-1}\right\}$ is quantified as $M=\sum_{s=0}^{z-1}f_s=2^n$ denominator, where $f_s$, $n\in\mathbb{N}$. For a symbol $s\in \Sigma$, rANS merges $s$ into a range of size $f_s$. Let $cdf_s=f_0+f_1+\cdots+f_{s-1}$ denote the cumulative frequency counts, and we define  $s\left(x\right)$ as symbol in $x\in \left [0,M \right )$ position, expressed as:
\begin{equation}
\overline{s}\left(x\right)=s \left(|x|_M\right), \qquad \text{where}\  s\left(x\right)=\text{min}\left\{ s:x<\sum_{i=0}^{s}f_i\right\}. 
\end{equation}
The encoding and decoding steps are as follows.
\begin{equation}\label{enc_ans}
\mathcal{C}(s,x)=M\lfloor x/f_s \rfloor+cdf_s+|x|_{f_s},
\end{equation}
\begin{equation}\label{dec_ans}
\mathcal{D}(x)=(s,f_s\lfloor x/M \rfloor+|x|_M-cdf_s), \qquad   \text{where}\  s=\overline{s}\left(x\right).
\end{equation}

As encoding continues, the value of state $x$ will eventually grow to infinity. Therefore, a streaming ANS~\cite{duda2013asymmetric} that enforces $x$ to remain in a specific interval $I$ is presented. Figure \ref{rANS} shows the details.

\section{Algorithms}\label{sec:3}
In this section, we first propose a BEMR with renormalization. Second, we give a variant of rANS coding. Then, the synchronization of encoder and decoder is shown. Finally, the benefits of the proposed scheme are discussed.

\subsection{BEMR with renormalization}\label{sec:3.1}
Given a number $x$ and a set of bases $\left\{ b_i\in \mathbb{N}^+ \right\}_{i=1}^N$, MRC converts $x$ to a mixed-radix representation $\left \langle r_1, r_{2},\cdots ,r_N \right \rangle$, and each $0\leq r_i<b_i$, via
\begin{equation}\label{eq:1}
	 r_i\gets |x|_ {b_i},  \quad x'\gets \lfloor x /b_i \rfloor.
\end{equation}
It can be seen that $x'<x$ as the conversion proceeds.

Clearly, when $x$ is big, it is expensive to perform MRC. To solve this issue, we perform renormalization on MRC to reduce the computational cost. The renormalization is also used in arithmetic coding and rANS. Analogously, in the proposed BEMR, we enforce $x$ to be within a certain range $I:=\left [L,H \right )$ by means of an auxiliary binary sequence $E$ stored in a deque $D$, where both $L$ and $H$ are natural numbers. During conversion, when $x<L$, we pop $t\in\mathbb{N}$ bits from deque $D$ and append it to the least significant digits of $x$ to increase its value before performing \eqref{eq:1}. Note that in the proposed BEMR, we always pop data from the front and push data from the back of deque.

Algorithm \ref{conversion} gives the details. In Algorithm \ref{conversion}, Lines $3$--$6$ maintain the state $x$ in $I$ by popping additional bits from deque and increase $x$ when it gets too small, where $T\in\mathbb{N}$. Lines $7$--$8$ convert a number $x$ to its mixed-radix representation. One can verify that the value of $x$ after each conversion is always within $I=\left [2^{T-t},2^T \right )$. Further, as the conversion is the exact opposite of the inverse conversion, the final $x$ needs to be transmitted along with the rest of the digit stream. Thus, the inverse converter knows what value $x$ to start with. For simplicity, given a binary sequence $E$, we define a function
\begin{equation}
	(r,x')\gets \mathrm{BEMR}_E^{T, t}\left(x, b\right)
\end{equation}
as the conversion of $x$ from binary to a base $b$, where $T$ and $t$ are optional positive integers, $r$ is the mixed-radix digit and $x'$ is the final $x$.

Next, we discuss the inverse BEMR (IBEMR) of the representation in base set $\left\{ b_i \right\}_{i=1}^N$ from a mixed-radix representation $\left\{ r_i \right\}_{i=1}^N$, where $r_i\in[0,b_i)$. Algorithm \ref{inverse-conversion} describes the proposed IBEMR. In Algorithm \ref{inverse-conversion}, Lines $4$--$8$ force $x$ to lie in $I$ by writing some bits from $x$ and decrease it when it gets too large. These written bits are exactly the bits extracted at Line $4$ in Algorithm \ref{conversion}. Finally, the data in deque $D$ is the desired binary representation $E'$. Let the function
\begin{equation}
	(E',x)\gets \mathrm{IBEMR}^{T, t}\left(x',b,r\right)	
\end{equation}	
denote the conversion of the base from $b$ to binary when the mixed-radix digit $r$ is given.

\begin{algorithm}[t]
	\caption{\label{conversion} Proposed BEMR}
	\begin{algorithmic}[1]
		\Require A binary sequence $E$, a set of bases $\left\{ b_i \right\}_{i=1}^N$ and a fixed integer $x$.
		\Ensure  A mixed-radix representation $\left\{ r_i \right\}_{i=1}^N$ and a final $x'$.
		\State Allocate a buffer $iobuf$
		\For{$i=N$ to $1$}
		\If{$x<b_i\times 2^{T-t}$}
		\State $Q\gets\mathrm{Deque}_E\left(t\right)$
		\State $x\gets (x\ll t)+Q$
		\EndIf
		\State $r_i\gets |x|_ {b_i}$
		\State $x\gets \lfloor x/b_i \rfloor$
		\State output $r_i$ to $iobuf$
		\EndFor
		\State $x'\gets x$\\
		\Return $iobuf$ and $x'$
	\end{algorithmic}
\end{algorithm}

\begin{algorithm}[t]
	\caption{\label{inverse-conversion} Proposed IBEMR}
	\begin{algorithmic}[1]
		\Require A set of bases $\left\{ b_i \right\}_{i=1}^N$, the mixed-radix representation $\left\{ r_i \right\}_{i=1}^N$ and a final $x'$.
		\Ensure A deque $D$ and an integer $x$.
		\State  Build a deque $D$
		\For{$i=1$ to $N$}
		\State $x'\gets b_i\times x'+r_i$
		\If{$x'\geq 2^{T}$}
		\State $mask\gets 2^t-1$
		\State $D.push\_back(x'\& mask)$
		\State $x'\gets x'\gg t$
		\EndIf
		\EndFor
		\State $x\gets x'$\\
		\Return $D$ and $x$.
	\end{algorithmic}
\end{algorithm}

\subsection{Variant of rANS coding}\label{sec:3.2}
\begin{algorithm}[t]
	\caption{\label{encoding} Proposed encoding algorithm}
	\begin{algorithmic}[1]
		\Require An input sequence $S$ of symbols and the appearances $\left\{ f_s \right\}_{s\in \Sigma}$ of symbols.
		\Ensure A deque $D$ and the final state $x$.
		\State Build a deque $D$
		\State $x\gets 2^T-1$
		\For{$i=N$ to $1$}
		\State $(r,x)\gets \mathrm{BEMR}_D^{T,vn}\left(x, f_{S[i]}\right)$
		\State $r\gets r+cdf_{S[i]}$
		\State $D.push\_back(r)$
		\EndFor\\
		\Return $D$ and $x$.
	\end{algorithmic}
\end{algorithm}

From \eqref{enc_ans}, one can see that $x$ becomes larger in the traditional rANS after encoding a symbol $s$. In contrast, in the proposed coding, we directly output $n$-bit $cdf_s+|x|_{f_s}$ for a symbol $s$, and decrease $x$ via $x\gets \lfloor x/f_s \rfloor$. Therefore, $x$ will become too small after encoding several symbols, eventually degrading the compression ratio. To solve this issue, the proposed rANS utilizes the BEMR described in Section~\ref{sec:3.1}. Precisely, BEMR forces the state $x$ always at a specific interval $I=\left [2^{T-vn},2^T \right )$, and $I_s:=\left [ f_s\times 2^{T-vn},2^T \right )$ denotes the interval corresponding to a symbol $s$, where $v\in\mathbb{N}$. Thus, for any state $x\in I_s$ in encoding, we have $\mathcal{C}\left(x,s\right)=\lfloor x/f_s \rfloor \in I$ and for $x\notin I_s$, we have $\mathcal{C}\left(x,s\right)\notin I$. For a symbol $s$, if the current $x$ is within $I_s$, we encode it to a $n$-bit digit and push the digit to a deque. Otherwise, we first pop data from the deque to enlarge $x$ before encoding.

As the encoded bit sequence is stored in a deque, the push and pop operations in the proposed encoding are possible at front or back of deque. The following gives an encoding algorithm in which data is pushed from the back and popped from the front. Specifically, for every input symbol $s$, if the current state $x\in I_s$, we directly push $n$-bit $cdf_s+|x|_ {f_s}$ to deque and reduce the value of $x$ by $x\gets \lfloor x/f_s \rfloor$. Otherwise, we first pop data from deque, appending them to the state $x$ to increase it. Notably, to ensure that there are enough bits in deque before renormalization, we initialize the state sufficiently large, such as $x = 2^T-1$. Algorithm \ref{encoding} presents the details. In Algorithm \ref{encoding}, if $x$ is a $64$-bit unsigned integer, we can choose $n=16, v=3, T=64-n=48$ to complete the encoding. Line $4$ pops $vn$ bits from deque for renormalization. Line $6$ outputs $n$ bits to deque each time. Finally, the remaining data in deque is the desired encoded bit sequence. 

\begin{algorithm}[t]
	\caption{\label{decoding} Proposed decoding algorithm}
	\begin{algorithmic}[1]
		\Require The final state $x$ of the encoder, the appearances $\left\{ f_s \right\}_{s\in \Sigma}$ of symbols and an encoded deque $D$.
		\Ensure Source symbol sequence $S$.
		\For{$i=1$ to $N$}
		\State $d\gets D.pop\_front()$
		\State $S[i]\gets \overline{s}(d)$
		\State $r\gets d-cdf_{S[i]}$
		\State $(D,x)\gets \mathrm{IBEMR}^{T,vn}\left(x, f_{S[i]},r\right)$
		\EndFor\\
		\Return $S$
	\end{algorithmic}
\end{algorithm}

\begin{figure}[t]
	\centering
	\includegraphics[width=0.7\linewidth]{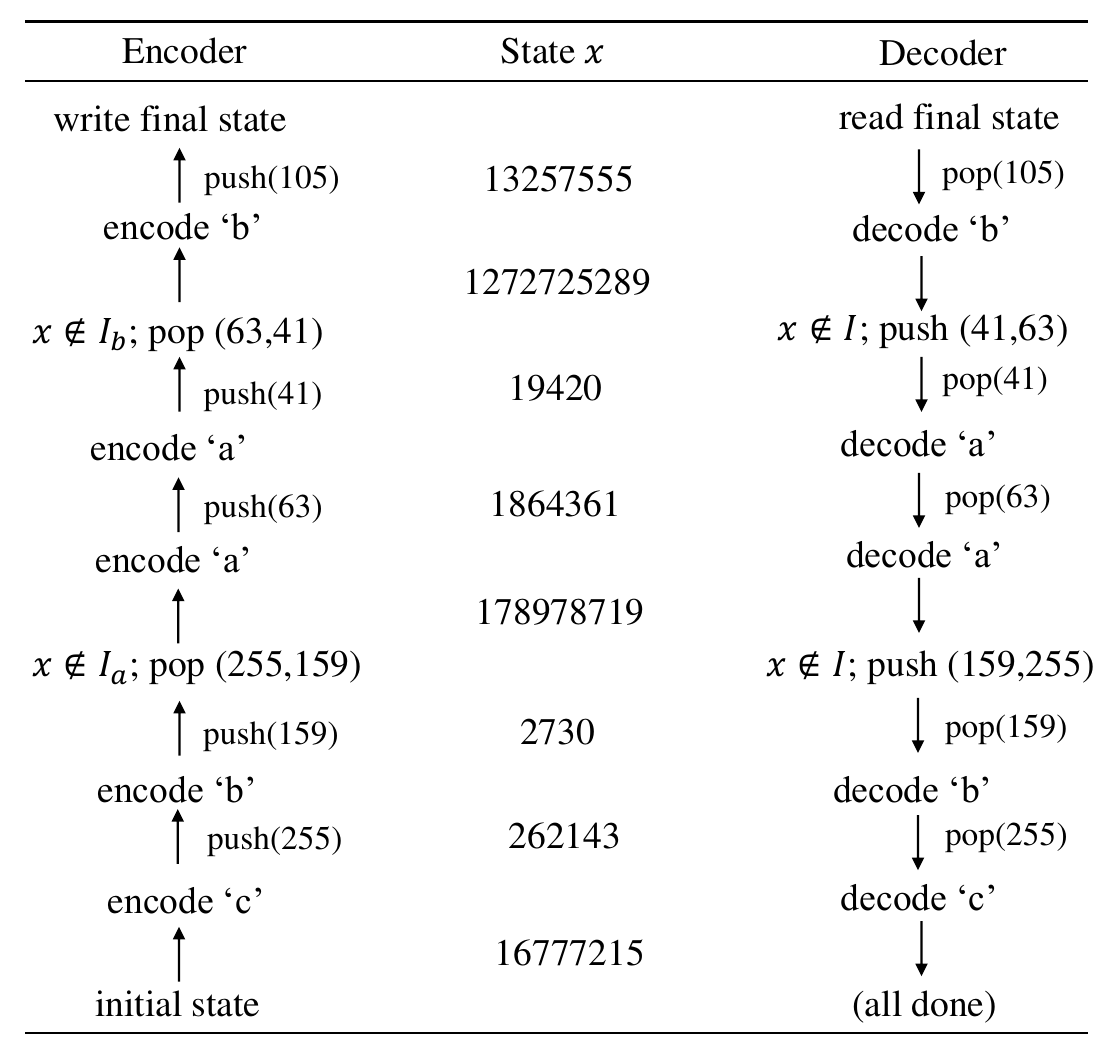}
	\caption{{Encoding example: Coding ``baabc" with $p(a) =p(b)=\frac{96}{256}$, $p(c) = \frac{64}{256}$.
			Encoder proceeds from bottom to top, decoder from top to bottom (as
			indicated). Both go through the exact same sequence of states and perform
			I/O in the same places.}}
	\label{fig1}
\end{figure}

Next, we discuss the proposed decoding procedure. First, we pop $n$-bit digit $d$ from deque and decode it for a symbol $s$. Then we update $x\gets f_{s}\times x+r$, where $r=d-cdf_{s}$. As more and more symbols are decoded from $x$, the value of $x$ becomes larger. When its value exceeds a given threshold $2^T$, we remove bits from $x$ to make it smaller. Algorithm \ref{decoding} describes the proposed decoding algorithm. In Algorithm \ref{decoding}, Lines $2$--$3$ decode the digit at the front of this deque to obtain a symbol. Line $5$ enlarges the value of $x$ and renormalizes it if necessary. Figure \ref{fig1} shows a worked-through example for the message ``baabc" with parameters $n=8$, $v=2$ and $T=32-n=24$. It can be seen that the encoder and decoder go through the same state just in opposite order. 

Notably, since the proposed coding uses $n$-bit as the basic unit for popping and pushing data from deque, the encoded bit sequence is $n$-bit aligned. For each $n$-bit digit $d$, we can use operation $\overline{s}(d)$ to decode a symbol. Therefore, $c$ $n$-bit can be decoded in parallel to obtain $c$ symbols, where $c\in\mathbb{N}$.

\subsection{Analysis}\label{sec:3.3}
In this subsection, we prove that the encoder and decoder go through the same state just in opposite order in the proposed coding. Let $X_{i}$ denote the state after encoding the  $(N+1-i)$-th symbol $s_i:=S[N+1-i]$. That is, $X_{1}$ is the state after encoding symbol $S[N]$. Accordingly, let $X'_{i-1}$ represent the state before encoding symbol $s_i$. From Section \ref{sec:3.2}, we have
\begin{equation} \label{*}
	X'_{i-1}
	=\left\{
	\begin{array}{ll}
		X_{i-1} & {\text{if } X_{i-1}\in \left [f_{s_i}\times2^{T-vn},2^T \right ),} \\ 
		(X_{i-1}\ll vn)+V & {\text {if } X_{i-1}\in \left [2^{T-vn},f_{s_i}\times2^{T-vn} \right ),}
	\end{array}\right. 
\end{equation}
and $X_i=\mathcal{C}(X'_{i-1},s_i)=\lfloor \frac{X'_{i-1}}{f_{s_i}}\rfloor$. Besides, we push $ R_{i}\triangleq cdf_{s_i}+|X'_{i-1}|_ {f_{s_i}}$ to deque.

In order to show that the encoder and decoder are always synchronized, it is necessary to prove that when state $X_i$ and $R_{i}$ are given, we can decode a symbol $s_i$ and $\mathcal{D}_1(X_i,s_i)=X'_{i-1}$, $\mathcal{D}_2(X'_{i-1})=X_{i-1}$, where $\mathcal{D}_1$ and $\mathcal{D}_2$ correspond to Line $3$ and Lines $4$--$8$ in Algorithm~\ref{inverse-conversion}, respectively. The proof is as follows. During decoding, we first pop $R_{i}$ from deque to decode a symbol $s_i$, then we have 
 	\begin{align*}
 	\mathcal{D}_1(X_i,s_i) &= f_{s_i}\times X_i+\left(R_{i}-cdf_{s_i} \right)\\
 	&=f_{s_i}\times \lfloor \frac{X'_{i-1}}{f_{s_i}}\rfloor+ |X'_{i-1}|_ {f_{s_i}}\\
 	&=X'_{i-1}.\\
 \end{align*}

From \eqref{*}, one can see that $X'_{i-1}\geq 2^T$ if and only if $X_{i-1}\in \left [2^{T-vn},f_{s_i}\times2^{T-vn} \right )$, and $X'_{i-1}< 2^T$ if and only if $X_{i-1}\in \left [f_{s_i}\times2^{T-vn},2^T \right )$. Next, there are two cases to calculate $\mathcal{D}_2(X'_{i-1})$ as follows.
\begin{enumerate}
	\item When $X'_{i-1}<2^T$, we have $\mathcal{D}_2(X'_{i-1})=X'_{i-1}=X_{i-1}$.
	\item When $X'_{i-1}\geq 2^T$, we have $\mathcal{D}_2(X'_{i-1})=\left(\left(X_{i-1}\ll vn \right)+V-V \right)\gg vn=X_{i-1}$.
\end{enumerate}
This completes the proof.

\subsection{Discussions}\label{sec:3.4}
In this subsection, we first show two features of the proposed coding when we pop data from the front and push data from the back of deque. Second, we discuss the differences between the traditional rANS and the proposed version.

\begin{figure}[t]
	\centering
	\includegraphics[width=0.8\linewidth]{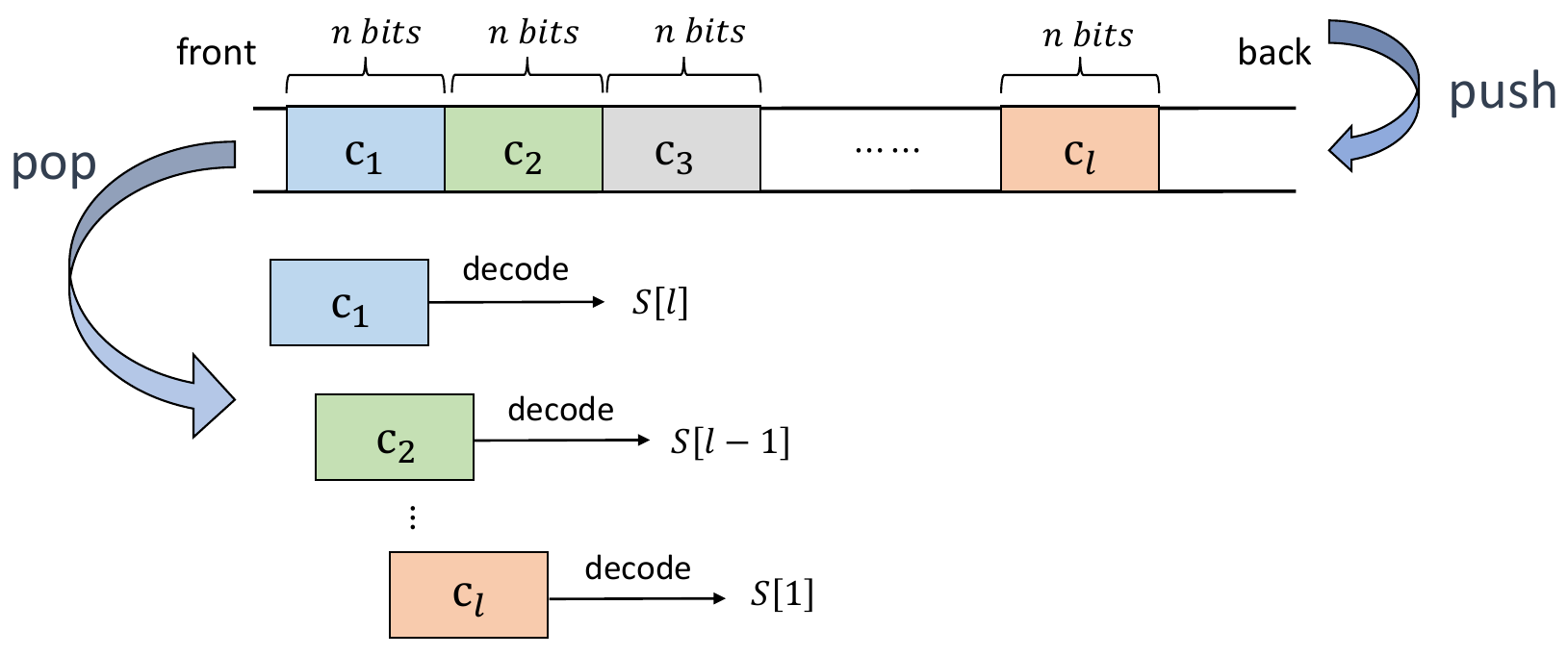}
	\caption{Diagram of direct access.
	}  
	\label{access}   
\end{figure}

\textbf{Partial direct access.}
As shown in Figure~\ref{access}, $\left\{ c_i \right\}_{i=1}^l$, where each $c_i$ is an $n$-bit digit, is the encoded stream in the deque after encoding. Then, we can directly decode $c_{l+1-j}$ to obtain the $j$-th symbol $S[j]$ in the source sequence, where $1\leq j\leq l$. This allows us to retrieve a symbol directly without decoding the whole encoded stream.

\textbf{Robustness.}
According to the above description, the final $l$ $n$-bit digits can directly decode the last encoded $l$ symbols. Therefore, if a small piece of data is altered in the compressed stream, the proposed coding ensures that several symbols after the altered position can still be correctly decoded. In contrast, for the traditional rANS, the altered bits in the encoded stream may cause all decoded symbols after the altered position to be incorrect.

\textbf{Comparisons between traditional rANS and the proposed algorithm.}
In Figure~\ref{rANS}, when the $b$-bit output in each renormalization is equal to the $n$-bit size of the frequency table, the encoded bit sequence of traditional rANS is $n$-bit aligned as well. This is equivalent to the proposed algorithm when $v = 1$, and the push/pop is at the back of deque. However, in other cases, that is, when $v > 1$ or the push/pop is at the front, the proposed algorithm is different to traditional rANS. Further, as the proposed coding outputs $vn$ bits in renormalization, this avoids poor performance when $n$ is small.

\section{Experiments}\label{sec:4}
In this section, we show the simulations of the proposed coding, traditional rANS (Base) and the interleaved ANS~\cite{giesen2014interleaved}. As this paper focuses on the performance of vanilla algorithms, the simulation does not consider SIMD implementations. The programs are written in C, compiled with GCC 7.4.0 with optimization level -O3. All programs are tested on the platform equipped with Intel(R) Core(TM) i7-6700K CPU @ 4.00GHz and 8 GB main memory on Ubuntu 18.04. The dataset used in the experiments is chosen from Calgary corpus\footnote{\url{http://www.data-compression.info/Corpora/CalgaryCorpus/}}. Table \ref{tab1} tabulates the information of the dataset, where the first two columns show the file name and alphabet size, respectively. The last column gives the total size of the input file in bytes.

\begin{table}[t]
	\begin{minipage}{0.48\linewidth}
		\centering
		\caption{Description of the dataset used}
		\label{tab1}
		\begin{tabular}{ccc}
			\hline 
			File &$|\Sigma|$& N \\
			\hline 
			book1  &82 &768771  \\
			book2  &96 &610856 \\
			paper1 &95 &53161  \\
			paper2 &91 &82199  \\
			news   &98 &377109 \\
			pic  &159 & 513216 \\
			geo  &256 & 102400\\
			obj1  &256 & 21504 \\	
			\hline 
		\end{tabular}
	
	\end{minipage}\begin{minipage}{0.48\linewidth}  
		\centering
		\caption{Compression ratio test}
		\label{ratio}
		\begin{tabular}{c|c|c}
			\hline 
			File & Base & Ours \\
			\hline
			book1 & 1.765 & 1.766\\
			book2 & 1.667 & 1.669\\
			paper1 & 1.603 & 1.605\\
			paper2 & 1.736 & 1.737\\
			news & 1.540 & 1.541\\
			pic & 6.605 & 6.593\\
			geo & 1.415 & 1.416\\
			obj1 & 1.343 & 1.344\\
			\hline
		\end{tabular}
	\end{minipage}
\end{table}

\begin{table}[H]
	\centering  
	\caption{Decompression throughput evaluation (MB/S)}
	\label{throughput}
	\begin{tabular}{c|c|c|c|c}
		\hline 
		File & \;Base & Ours & \;speed-up & 2-way interleaving\\
		\hline
		book1 & 167.5 & 353.7 & 2.11&290.9\\
		book2 & 165.4 & 334.4 & 2.02&288.7\\
		paper1 & 144.6 & 292.7 & 2.02&274.7\\
		paper2 & 149.2 & 348.8 & 2.34&252.9\\
		news & 161.5 & 313.6 & 1.94&274.3\\
		pic & 192.3 &  367.7& 1.91&316.0\\
		geo & 158.7& 308.2 & 1.94&251.3\\
		obj1 & 113.5 & 312.0 & 2.74&276.0\\
		\hline
	\end{tabular}
\end{table}

Next, we give two simulations and the details are as follows. First, we test the compression ratios of both coding schemes. The compression ratio is defined as
uncompressed size divided by compressed size in bytes, with higher ratios indicating stronger compression. Table \ref{ratio} shows the results. One can see that the proposed scheme is comparable in terms of compression ratio with the baseline. Second, we show the decompression throughput on various test files. Table \ref{throughput} tabulates the results, where 2-way interleaving denotes the interleaved ANS implementation, which uses two encoders with distinct states and writes to the same buffer. The throughput is defined as
\[
\text{Throughput}=\frac{\text{Size of input data  $\left(\text{MB}\right)$}}{\text{Time consumed  $\left(\text{Second}\right)$}},
\]
where MB stands for a megabyte. It can be seen that the proposed coding has a higher throughput with the speed-up factor about 2$\times$ compared to the baseline. Further, the decompression performance of the proposal in serial mode outperforms that of 2-way interleaving. This highlights the advantages of our proposal.

\section{Conclusions and future works}\label{sec:5}
In this paper, a BEMR is proposed by applying renormalization on radix conversion. Then based on BEMR, we present a variant of rANS coding. The simulations show that the proposed scheme in serial mode has a higher throughput than the baseline (with the speed-up factor about 2$\times$) without loss of compression ratio, and it outperforms the existing 2-way interleaving implementation. In addition, the synchronization of encoder and decoder is analyzed. 

In the future work, we aim to apply the proposed coding scheme to the time series compression, which focuses on encoding data in a much more compact format that
saves storage space without losing data. As the notable feature of time series data is that there is distinctive hot and cold data access, and recently written data is accessed more frequently. Based on this feature, we can encode the cold data first and then the hot data, which can greatly improve the overall query efficiency.

\section{References}
\bibliographystyle{IEEEbib}
\bibliography{refs}

\end{document}